\documentclass[reprint,aps,prl,showpacs,preprintnumbers,amsmath,amssymb]{revtex4-1}

\usepackage{graphicx}
\usepackage{dcolumn}
\usepackage{bm}
\usepackage{color}
\usepackage[normalem]{ulem} 

\begin{document}

\title{Experimental Constraint on an Exotic Spin- and Velocity-Dependent Interaction in the Sub-meV Range of Axion Mass with a Spin-Exchange Relaxation-Free Magnetometer}

\author{Young Jin Kim}
\email[Email address: ]{youngjin@lanl.gov}
\author{Ping-Han Chu}
\email[Email address: ]{pchu@lanl.gov}
\author{Igor Savukov}
\affiliation{Los Alamos National Laboratory, Los Alamos, New Mexico 87545, USA}


\begin{abstract}
We conducted a search for an exotic spin- and velocity-dependent interaction for polarized electrons with an experimental approach based on a high-sensitivity spin-exchange relaxation-free (SERF) magnetometer, which serves as both a source of polarized electrons and a magnetic-field sensor. The experiment aims to sensitively detect magnetic-fieldlike effects from the exotic interaction between the polarized electrons in a SERF vapor cell and unpolarized nucleons of a closely located solid-state mass. We report experimental results on the interaction with 82 h of data averaging, which sets an experimental limit on the coupling strength around $10^{-19}$ for the axion mass $m_a \lesssim 10^{-3}$ eV, within the important axion window.

\end{abstract}

                        
\maketitle
The extremely small value of the electric dipole moment of the neutron~\cite{Baker:2006,Pendlebury:2015} suggested the existence of new hypothetical fundamental bosons~\cite{Peccei,Weinberg,Wilczek} to resolve the strong $CP$ problem in the quantum chromodynamics, such as spin-0 axions~\cite{Peccei}. Several new theories resolving the problems of dark matter~\cite{Bertone}, dark energy~\cite{Joyce:2014kja}, and the hierarchy problem~\cite{PhysRevLett.115.221801} also require new bosons such as spin-0 axionlike particles (ALPs)~\cite{Jaeckel,Olive} and spin-1 dark photons~\cite{Essig}.
These bosons are predicted to mediate interactions between ordinary particles, such as photons, electrons, and nucleons~\cite{Jaeckel}. 
Current experiments for axion searches mainly focus on the axion coupling to the photon such as the the axion dark matter experiment (ADMX) using a resonant cavity~\cite{ADMX}, the CERN Axion Solar Telescope~\cite{Anastassopoulos:2017ftl}, and light shining-through-a-wall
experiments such as the any light
particle search~\cite{EHRET2010149}. The Particle Data Group has a review of recent efforts in this field~\cite{Patrignani:2016xqp}. 

In addition to the axion-photon coupling searches, recently exotic spin-dependent interactions associated with axions attracted new attention. The exotic spin-dependent interactions were introduced by Moody and Wilczek~\cite{PhysRevD.30.130} and later extended by Dobrescu and Mocioiu~\cite{Dobrescu}. A typical search for spin-dependent interactions requires a sensitive detector such as a torsion pendulum or an atomic magnetometer to measure an effective interaction similar to gravity or magnetism. A mass brought close to the detector can induce a new force if axions mediate the interaction between the mass and the detector. Therefore the searches for exotic spin-dependent interactions rely on a local supply of axions from a closely located mass and do not depend on cosmological and astrophysical axion sources. A recent review~\cite{Safronova:2017xyt} has described the theoretical motivation and experimental results of exotic spin-dependent interactions.

There are 15 possible exotic interactions between ordinary particles that contain static spin-dependent operators or both spin- and velocity-dependent operators~\cite{PhysRevD.30.130,Dobrescu}. Some of the interactions are not invariant under parity ($P$) or time-reversal ($T$) symmetries~\cite{Chu}; therefore their observation would provide new sources for $P$ and $T$ symmetry violations, which are essential for the matter-antimatter asymmetry of the Universe that cannot be explained by the standard model of particle physics~\cite{Canetti:2012zc}. To explore all the 15 interactions, we recently proposed an experimental approach based on a spin-exchange relaxation-free (SERF) atomic magnetometer~\cite{Chu}, the most sensitive cryogen-free magnetic-field sensor reaching femtotesla sensitivity~\cite{Kominis}. Unlike existing experiments, the SERF magnetometer in this approach serves as both a source of polarized electrons and a high-sensitivity detector, which leads to a simple, tabletop experimental design. This approach studies the exotic spin-dependent interactions between optically polarized electrons in a SERF vapor cell and atoms from an external solid-state mass~\cite{Chu}. 

Many experiments have been conducted for static spin-dependent interactions~\cite{Hammond,Hoedl,Terrano,Ni,Safronova:2017xyt} while spin- and velocity-dependent interactions have not been well investigated. In this Letter, we focus on the spin- and velocity-dependent interaction for electrons, adopting the numbering schemes in~\cite{Dobrescu,Leslie} in SI units, written as
\begin{align}
&V_{4+5} = -f_{4+5}\frac{\hbar^{2}}{8\pi m_{e}c}\left[\hat{\sigma_i}\cdot(\vec{v}\times\hat{r})\right]\left(\frac{1}{\lambda r}+\frac{1}{r^{2}}\right)e^{-r/\lambda},\label{eq:v45}
\end{align}	
where $\hbar$ is Planck's constant, $m_e$ is the mass of the polarized electron, $c$ is the speed of light in vacuum, $\hat{\sigma}_i$ is the $i$th spin vector of the polarized electron with $\vec{\sigma_i}=\hbar\hat{\sigma_i}/2$, $\hat{r}=\vec{r}/r$ is a unit vector in the direction between the polarized electrons and unpolarized nucleons, $\vec{v}$ is their relative velocity vector, and  $\lambda=\hbar/m_{\text{a}}c$ is the interaction range (the axion Compton wavelength), with $m_{\text{a}}$ being the axion mass. Here $f_{4+5}$ is the coupling strength constant for the interaction $V_{4+5}$, the combination of the scalar electron coupling with the scalar nucleon coupling~\cite{Leslie}. Apart from common interests on the interactions described in Ref.~\cite{Dobrescu}, a recent study showed that a modified electrodynamics can also generate spin- and velocity-dependent nonrelativistic potentials~\cite{PhysRevD.95.016006}. Recently, some new experimental results constraining spin- and velocity-dependent interactions for electrons such as torsion pendulums~\cite{PhysRevLett.97.021603}, helium fine-structure spectroscopy~\cite{PhysRevA.95.032505}, and antiprotonic helium spectroscopy~\cite{2018arXiv180100491F} were reported. A similar interaction for neutrons has been measured at the Paul Scherrer Institute~\cite{Piegsa:2012}. Here, we report an experimental constraint on the interaction $V_{4+5}$ between SERF polarized electrons and unpolarized nucleons for the axion mass $m_a \lesssim 10^{-3}$~eV, equivalent to the interaction range $\lambda \gtrsim 10^{-4}$~m, which is within the important axion window~\cite{Turner:1990}. 

Our experiment aimed to detect magnetic-fieldlike effects from the interaction. The interaction produces an effective magnetic field $\vec{A}_{4+5}$ at the location of the SERF vapor cell, which induces an energy shift of electrons in the SERF alkali-metal atoms $\Delta E$,
\begin{align}
V_{4+5}  =\gamma\hbar\hat{\sigma_i}\cdot\vec{A}_{4+5}= \Delta E\label{eq:deltaE}
\end{align}	
where $\gamma$ is the gyromagnetic ratio of the alkali atom, and  
\begin{align}
\vec{A}_{4+5} = -f_{4+5}\frac{\hbar}{8\pi m_{e}c\gamma}(\vec{v}\times\hat{r})\left(\frac{1}{\lambda r}+\frac{1}{r^{2}}\right)e^{-r/\lambda}.\label{eq:A}
\end{align}	
In a SERF magnetometer, a weak external magnetic field tilts the SERF polarized electron spins by a small angle proportional to the field's strength, which is measured with a probe laser beam~\cite{Karaulanov}. Similarly, the effective field $\vec{A}_{4+5}$ can tilt the SERF electron spins which can be sensitively detected in the SERF magnetometer~\cite{Chu}. 

\begin{figure}[t]
\centering
\includegraphics[width=3.45in]{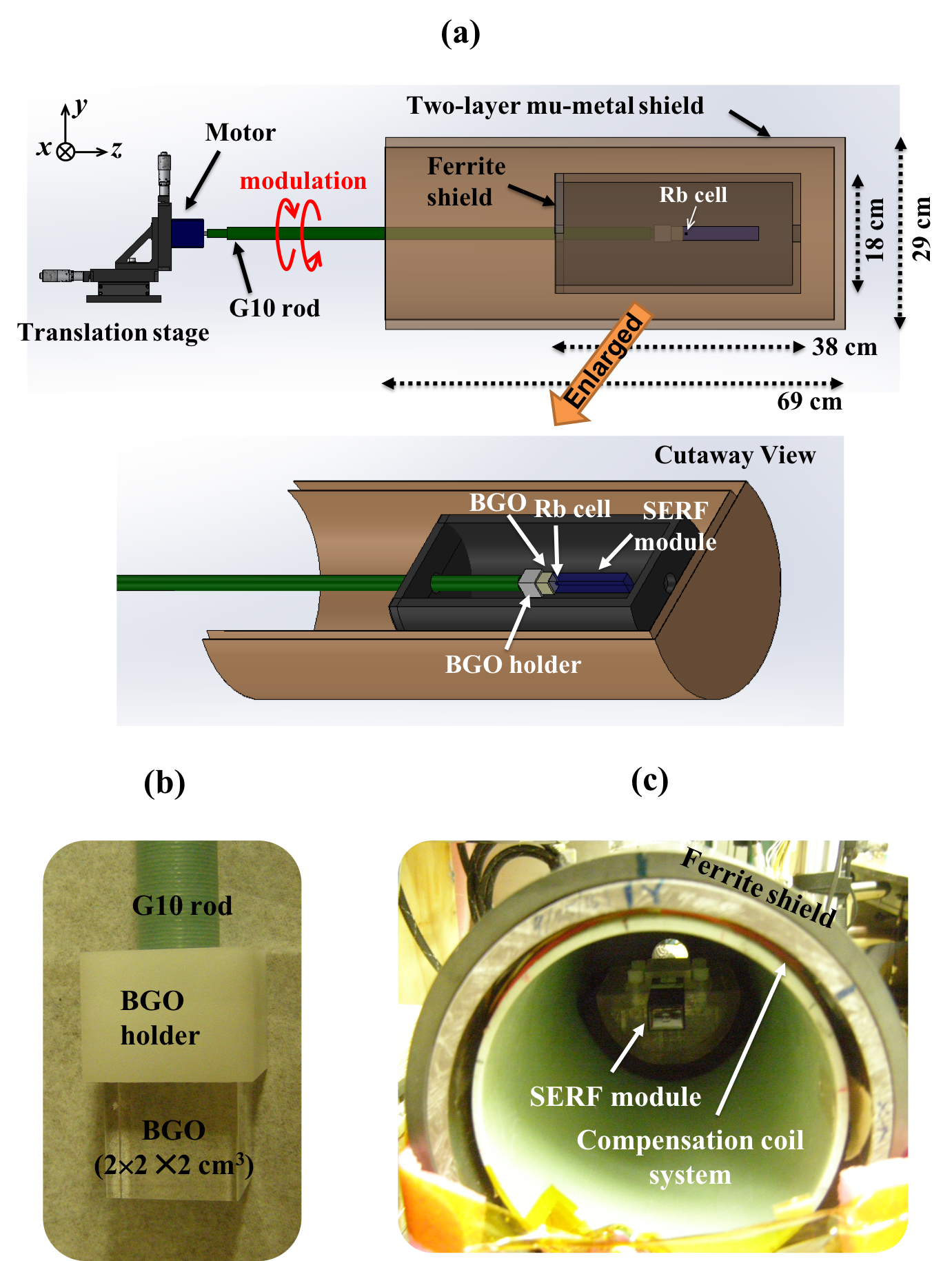}
\caption{\label{setup} (a) Side view of a schematic of the experimental setup to probe the exotic spin-dependent interaction $V_{4+5}$. An unpolarized BGO mass is placed next to a Rb vapor cell located inside the head of a SERF magnetometer module. The polarized Rb electron spins are oriented along the $y$ axis. The mass is rotated clockwise and counterclockwise around the $z$ axis to reduce systematic effects. (b) Photograph of the BGO mass connected to a G10 rod via a plastic holder to precisely control the position of the mass by using a three-axis translation stage. (c) Photograph of the SERF magnetometer module located inside a cylindrical ferrite shield (end cap not shown) that includes compensation coils to remove the residual field inside the shield.}
\end{figure}
The experimental setup to probe the interaction $V_{4+5}$ is shown in Fig.~\ref{setup}. For an unpolarized mass, we used a $2\times2\times2$~cm$^{3}$ cube-shaped nonmagnetic bismuth germanate insulator [Bi$_4$Ge$_3$O$_{12}$ (BGO)] with a high number density of nucleons ($4.3\times10^{24}$~cm$^{-3}$)~\footnote{Here we only consider the electron-nucleon interaction. But we can also consider unpolarized electrons in BGO mass, which will increase the total particle density roughly 50\% from the nucleon density.}, provided by Rexon Components, Inc. and shown in Fig.~\ref{setup}(b). We used a centimeter-scale SERF magnetometer, provided by QuSpin~\cite{quspin}, which contains a $3\times3\times3$~mm$^3$ $^{87}$Rb vapor cell with $\sim10^{13}$ Rb atoms and a single laser for both optical pumping and probing~\cite{Savukov}. This magnetometer is a compact, self-contained unit with all the necessary optical components that can be readily operated using the provided control software~\cite{Savukov}. The Rb spins were polarized along the $y$ axis and the magnetometer was sensitive to the magnetic field in the $z$ direction. The intrinsic field noise level of the magnetometer in the sensitive direction was measured to be 15~fT/Hz$^{1/2}$ at low frequencies between 5 and 100~Hz. In order to calibrate the magnetometer output voltage signals into magnetic-field signals, an internal coil mounted near the Rb cell generated a known calibration field. The bandwidth of the magnetometer was measured to be around 100~Hz. The magnetometer was surrounded with a cylindrical ferrite shield with end caps (18~cm diameter and 38~cm height), which was inserted into a two-layer open $\mu$-metal concentric cylindrical shield (26~cm inner diameter, 29~cm outer diameter, and 69~cm height) to suppress the effects of Earth's field, the external static fields, field gradients, and magnetic noise. The residual fields and linear field gradients inside the ferrite shield were suppressed by compensation coil systems [Fig.~\ref{setup}(c)].

The BGO mass, attached to a rigid G10 rod connected to a stepper motor (DMX-J-SA-17 provided by Arcus Technology) fixed on a three-axis translation stage (Thorlabs PT3), was positioned closely next to the magnetometer head by using the translation stage. The distance between the nearest vapor cell wall and the nearest part of the mass was set to $\sim$5~mm, which is limited by the position of the Rb vapor cell inside the SERF module. The centers of the Rb cell and the mass were aligned along the $z$ axis. The motor and the translation stage containing magnetic parts were placed outside the $\mu$-metal shield to reduce their effect on sensitive magnetic measurements. 

To create the relative velocity term for the interaction $V_{4+5}$, the BGO mass was rotated around the $z$ axis next to the SERF Rb cell using the motor. In this configuration, only the $z$ component of $\vec{A}_{4+5}$ remains, which tilts the polarized Rb electron spins by a small angle. The tilt is measured with the magnetometer's probe beam to nanoradian sensitivity. In order to cancel systematic effects, mainly due to trace magnetic contamination of the BGO mass ($7\times10^{-12}$~T) and the dc offset of the magnetometer (on the order of $10^{-10}$~T), we compared the magnetometer signals between clockwise and counterclockwise mass rotations. This works because the systematic effects are the same for the opposite rotations while the sign of the $\vec{A}_{4+5}$ is reversed due to only one velocity term in Eq.~(\ref{eq:A}): 
\begin{align}
\frac{1}{2}[(A_{4+5}+B_{\text{sys}})_\uparrow - (-A_{4+5}+B_{\text{sys}})_\downarrow]= A_{4+5}\label{eq:sys}
\end{align}	
where the symbols $\uparrow~(\downarrow)$ refer to the clockwise (counterclockwise) mass rotation, and $B_{\text{sys}}$ is the systematic effects. Furthermore, the SERF magnetometer and the shields were decoupled from the motor system rotating the mass, so that any mechanical vibration due to the mass motion is not observable in the magnetometer signals.

\begin{figure}[t]
\centering
\includegraphics[width=3.45in]{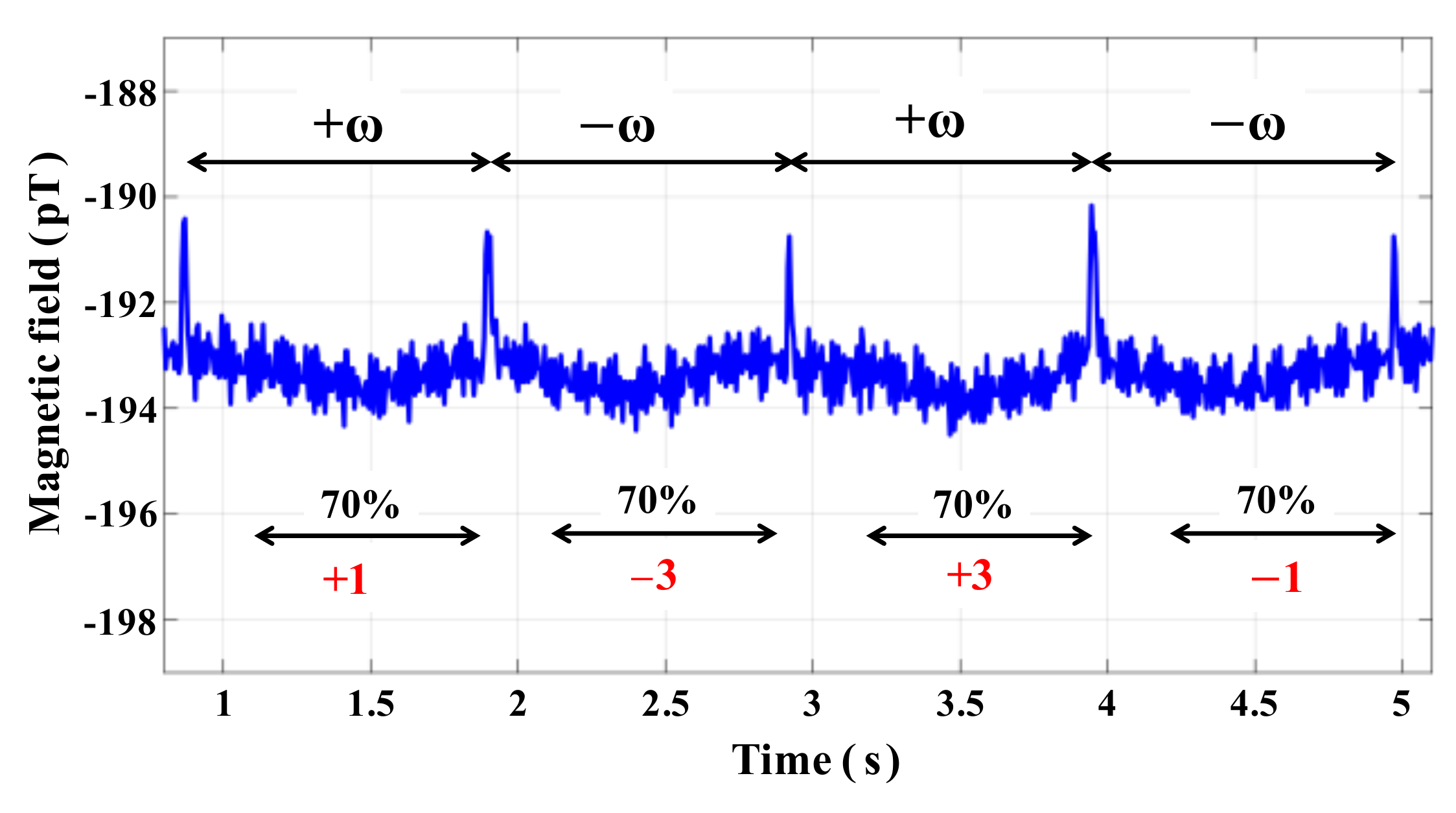}
\caption{\label{data} Time traces of SERF magnetometer signal showing two adjacent cycles of the mass rotation reversal. The BGO mass was rotated clockwise for 1~s and then rotated counterclockwise for 1~s. The angular velocity was $2\pi$~rad/s. The baseline did not change when the system started rotating. The spikes are associated with currents in the motor reversing the mass rotation. }
\end{figure}
The BGO mass was rotated clockwise for 1~s and then counterclockwise for 1~s at an angular velocity $\omega$ of $2\pi$~rad/s. This motion was continuously repeated and the SERF magnetometer signals were collected for 82 h. Figure~\ref{data} indicates a typical time trace of the magnetometer signal presenting two full cycles of the BGO mass rotation reversal. As the motor reversed the mass rotation, the magnetometer detected brief picotesla spikes associated with currents in the motor reversing the rotation direction. These spikes were chosen as the reference points for each half cycle. To extract the effective field $A_{4+5}$ we want to find the difference in the magnetometer signals between the two rotational states of the BGO mass [see Eq.~(\ref{eq:sys})]. However, inevitable slow drifts in the magnetometer signal, mainly due to the magnetometer electronics, will introduce a systematic bias to the data. We model the magnetometer signal as $B(t)=a+bt+ct^2\pm A_{4+5}$, where $a$ is a static offset, and $b$ and $c$ are coefficients of first- and second-order drift terms due to the magnetometer. Unlike the other terms, the target $A_{4+5}$ varies in sign with the rotational state of the BGO mass. To remove the $a, b, c$ terms we use a weighted sum of the data from each half cycle with a ``drift-correction algorithm." The process takes the weighted mean of the data from each half cycle using the weights [$+$1~$-$3~$+$3~$-$1]~\cite{kim}. The weights are chosen so that our algorithm removes not only the static field, but the magnetometer drift terms $b$ and $c$, eliminating systematic effects due to slow drifts in the magnetometer to second order:
\begin{align}
\Delta B & = B(\tau/2)-3B(\tau)+3B(3\tau/2)-B(2\tau)\notag
 \\
& = 8A_{4+5}.
\label{eq:drift}
\end{align}	
Here $\tau$ is the time period of the rotation reversal cycle. To obtain $A_{4+5}$, $\Delta B$ has to be divided by eight and to eliminate the effects from the spikes associated with the motor currents, only the last 70\% of the data in each half cycle were used, as indicated in Fig.~\ref{data}. As a numerical example, in the case of Fig.~\ref{data}, $\Delta B= -193.44~\text{pT}-3\times(-193.42~\text{pT})+3\times(-193.62~\text{pT})-(-193.42~\text{pT})=-0.62~\text{pT}$. To quantify the systematic effect due to the slow drifts, we extracted $A_{4+5}$ without applying the drift-correction algorithm by taking the mean of the last 70\% of the data in each half cycle and finding the difference in the mean values between the two rotational states of the BGO mass [see Eq.~(\ref{eq:sys})]. The systematic effect was measured to be $2\times10^{-15}$~T, which corresponds to a upper bound of the coupling strength $f_{4+5}$.

\begin{figure}[t]
\centering
\includegraphics[width=3.5in]{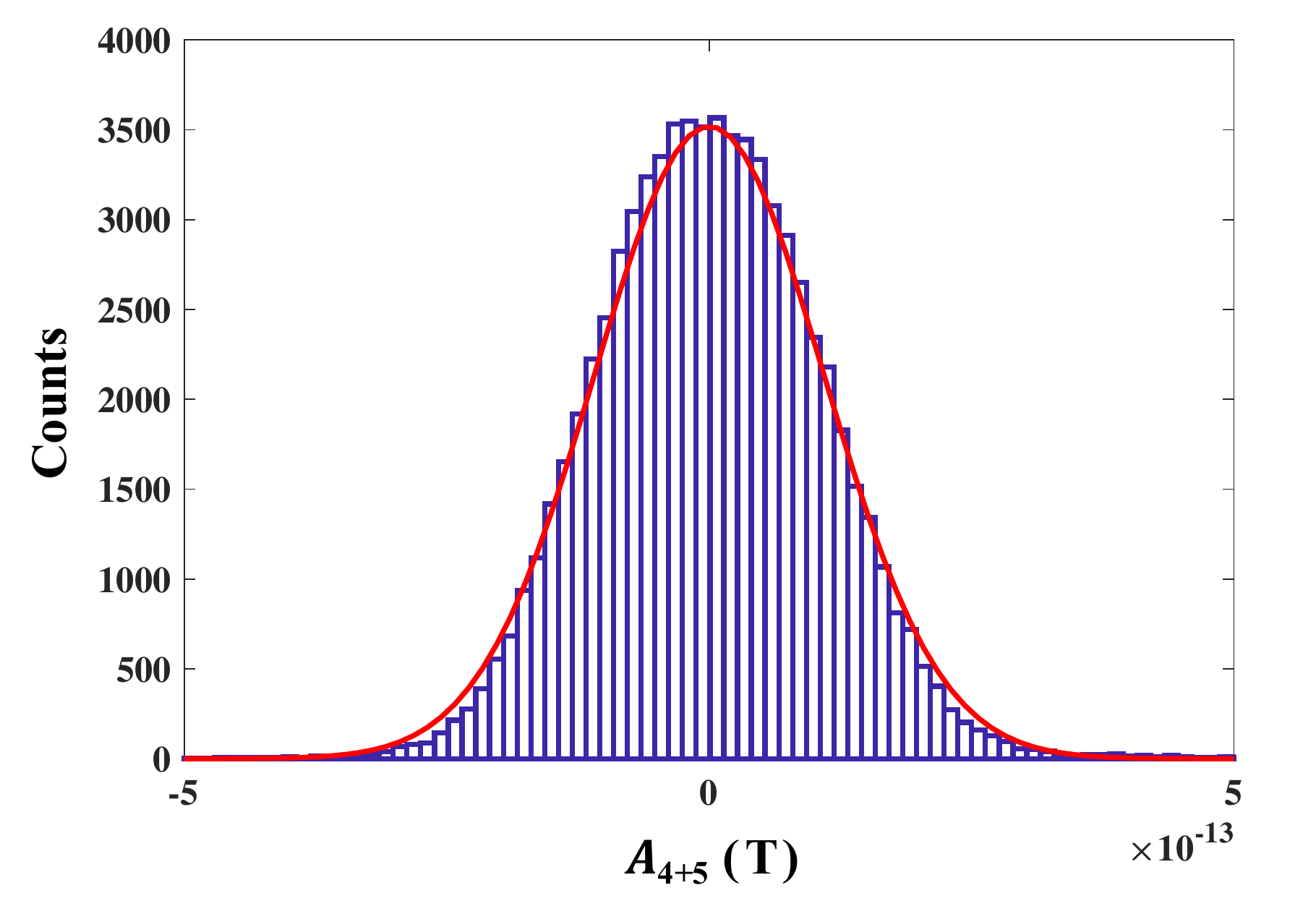}
\caption{\label{hist} Distribution of the effective field $A_{4+5}$ processed with the drift-correction algorithm. The data were collected for 82 h. The solid line indicates a fit to a Gaussian distribution. The slight deviation from the Gaussian distribution is mainly due to slow drifts higher than second order in the magnetometer signal, which were not canceled by the drift-correction algorithm.}
\end{figure}
Figure~\ref{hist} shows a histogram of $A_{4+5}$ values obtained with the drift-correction algorithm from data collected for 82 h. The histogram was fit to a Gaussian distribution, giving $A_{4+5}=(1.27\pm4.02)\times10^{-16}$~T or  $(2.31\pm7.30)\times10^{-20}$~eV in terms of the energy shift of Rb atoms with the gyromagnetic ratio of $2\pi\times7.0\times10^9$~Hz/T~\cite{Karaulanov}. The dominant  systematic effects due to magnetic impurities buried inside the BGO mass have been effectively suppressed below the statistical sensitivity of $4.02\times10^{-16}$~T by subtracting opposite rotation signals.

To constrain the coupling strength $f_{4+5}$, we performed a Monte Carlo integration for the interaction potential~\cite{Chu}. We generated $2^{20}$ random point pairs inside the volumes of the BGO mass and the Rb vapor cell, and calculated the interaction potential for each pair using Eq.~(\ref{eq:v45}). For a given interaction range between 10$^{-1}$ and 10$^{-6}$~m, we summed and normalized the potential for the nucleon density of BGO mass. Only the force along the z axis survived. The potential can only affect and tilt the electron spin of Rb atoms along the z-axis which can be detected by the probing beam of the SERF system. As described in Eq.~(\ref{eq:deltaE}), the experimental limit to the coupling strength was derived by dividing the experimental sensitivity of the energy shift for Rb atoms in the above by the calculated  potential. The error of the Monte Carlo calculations is less than 1\%, which is sufficient for coupling strength estimates from our experiments.

\begin{figure}[t]
\centering
\includegraphics[width=3.4in]{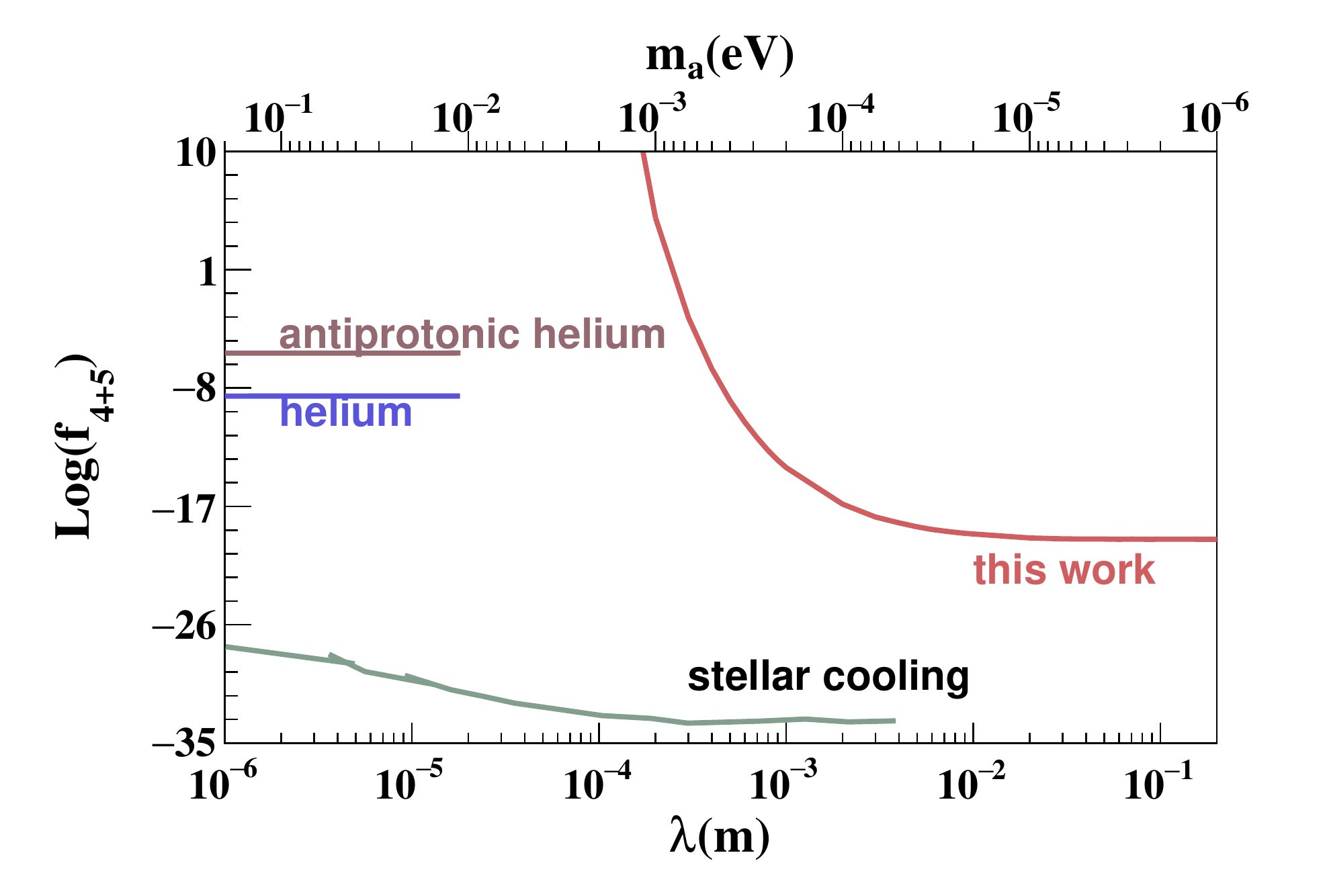}
\caption{\label{constraint} The red curve shows the experimental limit of this work on the interaction $V_{4+5}$ for $m_a \lesssim 10^{-3}$~eV and $\lambda \gtrsim 10^{-4}$~m between polarized SERF Rb electron spins and unpolarized BGO nucleons  as a function of the interaction range (bottom axis) and the axion mass (top axis) with the 82 h of data collection time. The coupling is the combination of the scalar electron coupling and the scalar nucleon coupling. The curve of stellar cooling is the combination of the scalar electron coupling derived from stellar cooling and the scalar nucleon coupling derived from short-range gravity experiments, discussed in detail in Refs.~\cite{Leslie, PhysRevD.86.015001}. Recent results from the measurement of helium fine-structure spectroscopy~\cite{PhysRevA.95.032505} of the scalar electron and scalar electron couplings, and antiprotonic helium spectroscopy~\cite{2018arXiv180100491F} of the scalar electron coupling and the scalar antiproton coupling are shown below the range of $10^{-5}$ m. }
\end{figure}
Figure~\ref{constraint} shows the experimentally set limit on the coupling strength of the interaction $V_{4+5}$ between unpolarized nucleons of the BGO mass and polarized Rb electron spins in the SERF vapor cell in the interaction range above $10^{-4}$~m, with the experimental sensitivity of $4.02\times10^{-16}$~T. This implies that our experiment is sensitive in the axion mass range below $10^{-3}$~eV. Unlike the axion experiments using cavities such as ADMX, the SERF magnetometer can simultaneously scan the axion mass range without tuning parameters for each specific axion mass.
The interaction $V_{4+5}$ for polarized electrons is experimentally constrained in this mass range for the first time, opening up new ranges of searches for the exotic spin-dependent interactions.

In conclusion, we searched for an exotic spin- and velocity-dependent interaction for polarized electrons using an experimental method based on a SERF magnetometer. We reported the experimental limit on the interaction, free of systematic signals, in the interaction range of $10^{-1}$ -- $10^{-4}$~m corresponding to the axion mass of $10^{-6}$ -- $10^{-3}$~eV. Although no signal from axions for $V_{4+5}$ was detected, we plan to probe the other possible interactions~\cite{Chu}, $V_{12+13}$ and $V_{9+10}$, between the SERF polarized electrons and the nucleons of the unpolarized BGO mass by properly moving the mass next to the vapor cell. Torsion balance experiments have set constraints on the interaction $V_{12+13}$ for the axial electron coupling and the vector nucleon coupling at the interaction range $\lambda> 10^8$~m~\cite{Adelberger:2009}. To the best of our knowledge, this interaction has never been experimentally constrained at the range of  $10^{-3}$~m within the axion window; therefore our experiments will shed light on the new direction of axion searches. 

The authors thank Pulak Nath and Shaun Newman for help in experiment. The authors gratefully acknowledge that this work was supported by the Los Alamos National Laboratory LDRD office through Grant No. 20180129ER.

\bibliography{reference}
\end{document}